\documentclass[twocolumn,showpacs,showkeys,reprint,amsmath,amssymb,superscriptaddress,prx]{revtex4-1}

\usepackage{hyperref}
\usepackage{graphicx}
\usepackage{amsmath}

\begin{document}

\title{Magnetism-induced topological transition in EuAs$_3$}

\author{Erjian Cheng}
\altaffiliation{These authors contributed equally to this work}
\affiliation{State Key Laboratory of Surface Physics, Department of Physics, and Laboratory of Advanced Materials, Fudan University, Shanghai 200433, China}

\author{Wei Xia}
\altaffiliation{These authors contributed equally to this work}
\affiliation{School of Physical Science and Technology, ShanghaiTech University, Shanghai 200031, China}
\affiliation{ShanghaiTech Laboratory for Topological Physics, Shanghai 201210, China}

\author{Xianbiao Shi}
\altaffiliation{These authors contributed equally to this work}
\affiliation{State Key Laboratory of Advanced Welding \&\ Joining and Flexible Printed Electronics Technology Center, Harbin Institute of Technology, Shenzhen 518055, China}
\affiliation{Key Laboratory of Micro-systems and Micro-structures Manufacturing of Ministry of Education, Harbin Institute of Technology, Harbin 150001, China}

\author{Chengwei Wang}
\altaffiliation{These authors contributed equally to this work}
\affiliation{School of Physical Science and Technology, ShanghaiTech University, Shanghai 200031, China}
\affiliation{State Key Laboratory of Functional Material for Informatics, Shanghai Institute of Microsystem and Information Technology, Chinese Academy of Sciences, Shanghai 200050, China}

\author{Chuanying Xi}
\affiliation{Anhui Province Key Laboratory of Condensed Matter Physics at Extreme Conditions, High Magnetic Field Laboratory of the Chinese Academy of Sciences, Hefei, Anhui 230031, China}

\author{Shaowen Xu}
\affiliation{Department of Physics, Shanghai University, Shanghai 200444, China}

\author{Darren C. Peets}
\affiliation{Ningbo Institute of Materials Technology and Engineering, Chinese Academy of Sciences, Ningbo, Zhejiang 315201, China}
\affiliation{Institute for Solid State and Materials Physics, Technical University of Dresden, 01062 Dresden, Germany}

\author{Linshu Wang}
\affiliation{State Key Laboratory of Surface Physics, Department of Physics, and Laboratory of Advanced Materials, Fudan University, Shanghai 200433, China}

\author{Hao Su}
\affiliation{School of Physical Science and Technology, ShanghaiTech University, Shanghai 200031, China}

\author{Li Pi}
\affiliation{Anhui Province Key Laboratory of Condensed Matter Physics at Extreme Conditions, High Magnetic Field Laboratory of the Chinese Academy of Sciences, Hefei, Anhui 230031, China}

\author{Wei Ren}
\affiliation{Department of Physics, Shanghai University, Shanghai 200444, China}

\author{Xia Wang}
\affiliation{School of Physical Science and Technology, ShanghaiTech University, Shanghai 200031, China}

\author{Na Yu}
\affiliation{School of Physical Science and Technology, ShanghaiTech University, Shanghai 200031, China}

\author{Yulin Chen}
\affiliation{School of Physical Science and Technology, ShanghaiTech University, Shanghai 200031, China}
\affiliation{ShanghaiTech Laboratory for Topological Physics, Shanghai 201210, China}
\affiliation{Department of Physics, University of Oxford, Oxford, OX1 3PU, United Kingdom}

\author{Weiwei Zhao}
\affiliation{State Key Laboratory of Advanced Welding \&\ Joining and Flexible Printed Electronics Technology Center, Harbin Institute of Technology, Shenzhen 518055, China}
\affiliation{Key Laboratory of Micro-systems and Micro-structures Manufacturing of Ministry of Education, Harbin Institute of Technology, Harbin 150001, China}

\author{Zhongkai Liu}
\email{liuzhk@shanghaitech.edu.cn}
\affiliation{School of Physical Science and Technology, ShanghaiTech University, Shanghai 200031, China}
\affiliation{ShanghaiTech Laboratory for Topological Physics, Shanghai 201210, China}

\author{Yanfeng Guo}
\email{guoyf@shanghaitech.edu.cn}
\affiliation{School of Physical Science and Technology, ShanghaiTech University, Shanghai 200031, China}

\author{Shiyan Li}
\email{shiyan$\_$li@fudan.edu.cn}
\affiliation{State Key Laboratory of Surface Physics, Department of Physics, and Laboratory of Advanced Materials, Fudan University, Shanghai 200433, China}
\affiliation{Collaborative Innovation Center of Advanced Microstructures, Nanjing 210093, China}

\begin{abstract}

The nature of the interaction between magnetism and topology in magnetic topological semimetals remains mysterious, but may be expected to lead to a variety of novel physics. We present $ab$ $initio$ band calculations, electrical transport and angle-resolved photoemission spectroscopy (ARPES) measurements on the magnetic semimetal EuAs$_3$, demonstrating a magnetism-induced topological transition from a topological nodal-line semimetal in the paramagnetic or the spin-polarized state to a topological massive Dirac metal in the antiferromagnetic (AFM) ground state at low temperature, featuring a pair of massive Dirac points, inverted bands and topological surface states on the (010) surface. Shubnikov-de Haas (SdH) oscillations in the AFM state identify nonzero Berry phase and a negative longitudinal magnetoresistance ($n$-LMR) induced by the chiral anomaly, confirming the topological nature predicted by band calculations. When magnetic moments are fully polarized by an external magnetic field, an unsaturated and extremely large magnetoresistance (XMR) of $\sim$ 2$\times10^5$\,\%\ at 1.8\,K and 28.3\,T is observed, likely arising from topological protection. Consistent with band calculations for the spin-polarized state, four new bands in quantum oscillations different from those in the AFM state are discerned, of which two are topologically protected. Nodal-line structures at the $Y$ point in the Brillouin zone (BZ) are proposed in both the spin-polarized and paramagnetic states, and the latter is proven by ARPES. Moreover, a temperature-induced Lifshitz transition accompanied by the emergence of a new band below 3\,K is revealed. These results indicate that magnetic EuAs$_3$ provides a rich platform to explore exotic physics arising from the interaction of magnetism with topology.

\end{abstract}

\maketitle

\section{Introduction}

\begin{figure*}
\includegraphics[clip,width=0.95\textwidth]{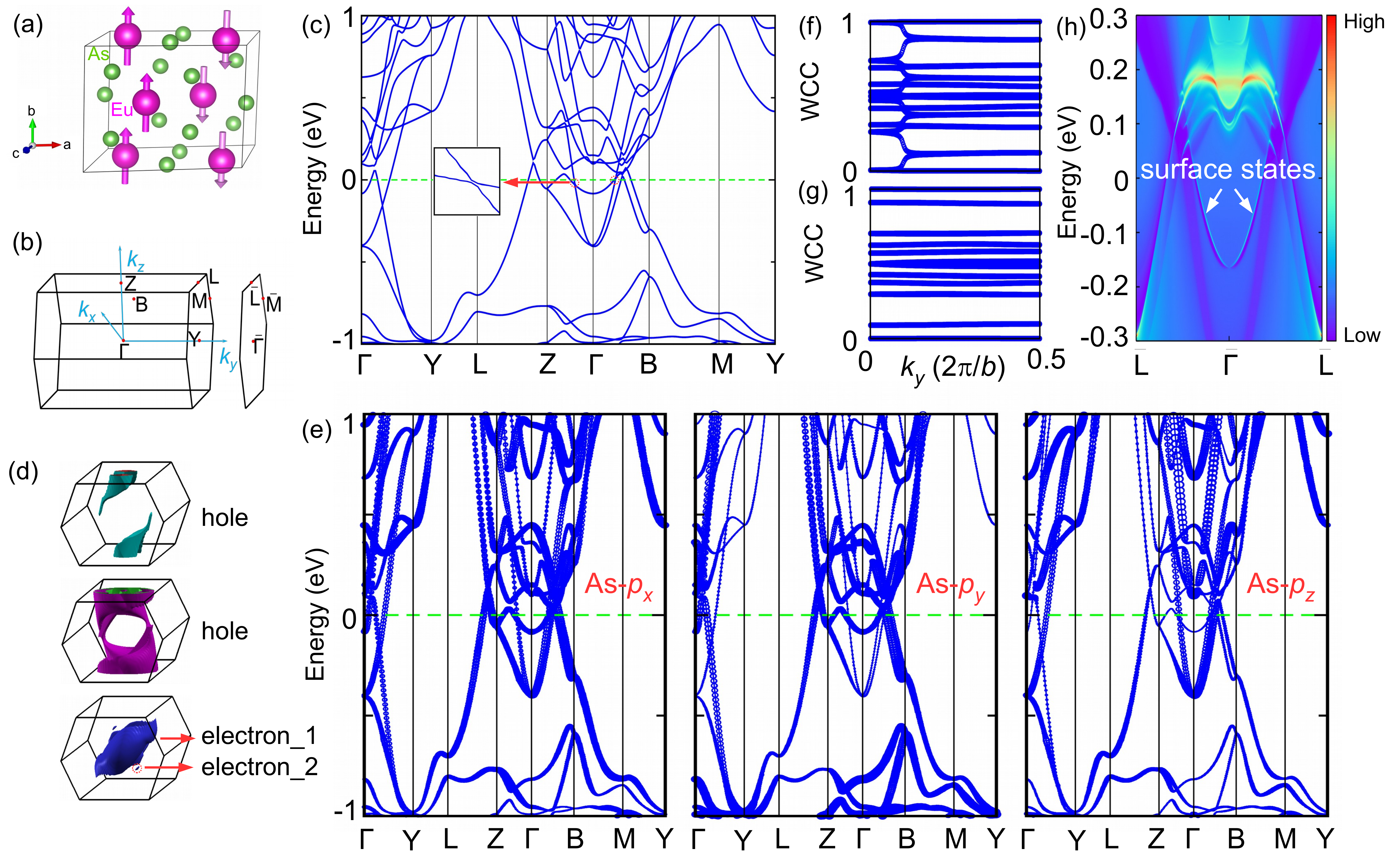}
\caption{\label{fig1}
Topological massive Dirac metal state in the antiferromagnetic state of EuAs$_3$, revealed by band calculations.
(a) Schematic for the crystal structure of EuAs$_3$ in the doubled magnetic unit cell. The arrows on Eu$^{2+}$ represent the spin directions, which are parallel and antiparallel to the $b$ axis. 
(b) Bulk and (110)-projected surface BZs of the doubled magnetic unit cell of EuAs$_3$ with several high-symmetry points marked. 
(c) Band structure of EuAs$_3$ from GGA+SOC+$U$ ($U$ = 5\,eV) calculations for the AFM ground state. The inset shows the massive Dirac point with a small gap. 
(d) Fermi surfaces of EuAs$_3$ derived from the band structure. 
(e) Projected band structure of EuAs$_3$, where the symbol size represents the projected weight of Bloch states onto the As $p_x$, $p_y$, and $p_z$ orbitals as labelled. Band inversion can be observed at the $\Gamma$ point. The Wannier charge center is calculated in the (f) $k_z$ = 0 and (g) $k_z$ = 0.5 planes. 
(h) Calculated surface states on the (010) surface. The nontrivial topological surface states are clearly visible.
}
\end{figure*}

Topological semimetals (TSMs) including Dirac, Weyl, nodal-line, and triple-point semimetals, can be divided into two categories---non-magnetic and magnetic TSMs---depending on whether magnetism is involved \cite{Armitage,Burkov,Xu}. Compared with better-known non-magnetic TSMs, magnetic TSMs have unique properties due to their broken time-reversal symmetry (TRS): for example, nonzero net Berry curvatures which can induce anomalous Hall or Nernst effects, only one pair of Weyl nodes for some magnetic Weyl semimetals, and a good ability to manipulate the spin for spintronics applications \cite{Xu}. Moreover, when magnetism is involved, interactions of the external magnetic field with the magnetic moments can result in exotic properties, such as Weyl states induced by magnetic exchange \cite{CeSb,EuCd2Sb2}. However, in contrast to non-magnetic TSMs, theoretical predictions and experimental studies on magnetic TSMs are rarer and more difficult due to the complexity of the magnetic configuration for calculations and the difficulty in synthesis of single crystals \cite{Xu}. In fact, the very nature of the interaction between magnetism and topology in magnetic TSMs remains mysterious. Given how few such compounds are known, seeking and fully characterizing new magnetic TSMs is a priority for the new light they may shed on these issues.

Recently, the non-magnetic CaP$_3$ family of materials was proposed as potential host of topological nodal-line (TNL) semimetals \cite{CaP3}, among which SrAs$_3$ possesses a TNL feature at ambient pressure \cite{SrAs3-1,SrAs3-2,SrAs3-3} and exotic properties under high pressure \cite{SrAs3-4}. Isostructural with SrAs$_3$, EuAs$_3$ orders in an incommensurate AFM state at $T_N$ = 11 K, and then undergoes an incommensurate-commensurate lock-in phase transition at $T_L$ = 10.3 K, producing a collinear AFM ground state \cite{EuAs3_1,EuAs3_2,EuAs3_3,EuAs3_4,EuAs3_5,EuAs3_6}. Previous electrical transport studies found an extremely anisotropic MR, which is strongly related to the magnetic configuration of EuAs$_3$ \cite{EuAs3-MR}. However, experiments sensitive to the topology have not been reported on EuAs$_3$.

In this paper, we demonstrate a magnetism-induced topological transition from a TNL semimetal in the paramagnetic or the spin-polarized state to a topological massive Dirac (TMD) metal in the AFM ground state. First, we explore the band structure in the AFM ground state through band calculations and transport measurements, demonstrating that EuAs$_3$ is a magnetic TMD metal. Second, SdH oscillations and band calculations in the spin-polarized state are displayed, yielding a proposal that EuAs$_3$ is a TNL semimetal with an XMR of $\sim$ $2\times10^5$\,\%\ at 1.8 K and 28.3 T. Third, our ARPES results in the paramagnetic state verify the nodal-line structure as predicted by band calculations. Ultimately, the origin of the XMR and a temperature-induced Lifshitz transition are revealed.

\begin{figure*}
\includegraphics[clip,width=0.85\textwidth]{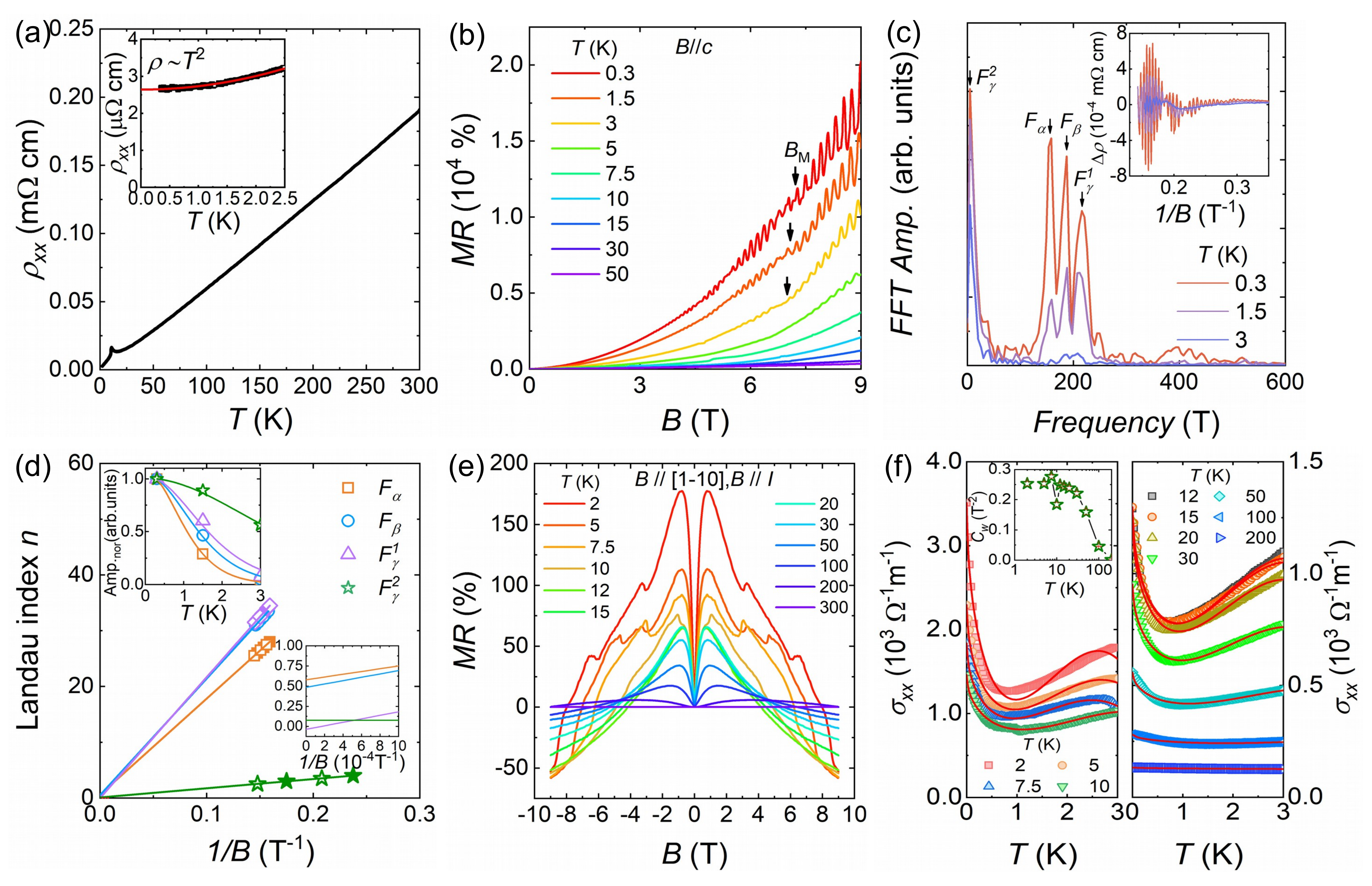}
\caption{\label{fig2}
Quantum oscillations and negative longitudinal magnetoresistance ($n$-LMR) in the antiferromagnetic state of EuAs$_3$.
(a) Resistivity of EuAs$_3$ single crystal in zero magnetic field. The inset shows the fit to the low-temperature data. 
(b) MR accompanied by distinct SdH oscillations. $B_M$ represents the critical magnetic field which induces a magnetic transition from a collinear antiferromagnetic phase to a polarized ferromagnetic phase. 
(c) FFT results at various temperatures. The inset displays the oscillatory component $\rho_{xx}$ below $B_M$. Four bands, i.e., $\alpha$, $\beta$, $\gamma^1$ and $\gamma^2$, can be distinguished. The latter two construct one electron sheet. 
(d) Landau index $n$ plotted against 1/$B$ for the SdH oscillations at 0.3 K. Lines represent linear fits. The right inset shows the extrapolation of 1/$B$ to zero. The left inset shows the normalized FFT amplitude (Amp.$_{nor}$) as a function of temperature, and the solid lines represent fits to the Lifshitz-Kosevich formula. 
(e) $n$-LMR measured with magnetic field parallel to the electric current $I$ at various temperatures. 
(f) Longitudinal conductivity at various temperatures fit to the Adler-Bell-Jackiw chiral anomaly equation. The inset shows the emergence of a positive parameter originating from the chiral anomaly $C_w$.
}
\end{figure*}

\section{Topological properties in the AFM state}

\subsection {Band claculations for the AFM state}

EuAs$_3$ crystallizes in a monoclinic structure (space group $C$2/$m$, No.\ 12), and the magnetic moments of Eu$^{2+}$ are oriented parallel and antiparallel to the monoclinic $b$ axis \cite{EuAs3_1,EuAs3_2,EuAs3_3,EuAs3_4,EuAs3_5,EuAs3_6}, as plotted in Fig.~\ref{fig1}(a). Methods for single-crystal growth, experiments, and band calculations can be found in the Supplementary Information. Figure~\ref{fig4}(c) shows the bulk BZ of EuAs$_3$ and Fig.~\ref{fig1}(b) shows the bulk and (110) surface BZs in the doubled unit cell corresponding to its antiferromagnetic ground state. The calculated band structure including spin-orbit coupling (SOC) in this magnetic ground state as determined by neutron diffraction experiments \cite{EuAs3_2} for EuAs$_3$ is displayed in Fig.~\ref{fig1}(c). In addition to topological bands around the $\Gamma$ point, several trivial bands cross the Fermi level, indicating that EuAs$_3$ is a metal rather than a semimetal. In magnetic systems, TRS is broken. To preserve the Dirac node, extra symmetries, for example the combination of inversion ($I$) and time-reversal ($T$) symmetries, i.e., $IT$, are necessary \cite{Xu}. The Dirac band crossing is not topologically protected, and it can be gapped out by SOC to turn into a gapped dispersion of massive Dirac fermions \cite{Xu}. Following this clue, two massive Dirac points around $\Gamma$ point are identified, as shown in the inset to Fig.~\ref{fig1}(c). The complicated Fermi surface is composed of two hole sheets and one electron sheet in the AFM state (Fig.~\ref{fig1}(d)), all of three-dimensional (3D) character. The electron sheet consists of two individual pockets, i.e., electron$\_$1 and electron$\_$2, which can both be detected by quantum oscillations.

Projected band structure analysis shows that the low-energy states near the Fermi level are dominated by As-4$p$ states (Fig.~\ref{fig1}(e)). There are clear signatures of band inversion between As-$p_{x,y}$ and As-$p_z$ orbitals at the $\Gamma$ point. To identify the topological character, we calculated the Z$_2$ invariant by employing the Willson loop method \cite{Willson_loop}. Figures~\ref{fig1}(f) and \ref{fig1}(g) show the evolution of the Wannier charge center (WCC) on two representative planes of the bulk BZ. From the calculations, the Z$_2$ invariant for the $k_z$ = 0 plane is 1, whereas Z$_2$ is 0 for the $k_z$ = 0.5 plane, providing strong evidence for nontrivial topology. Moreover, topologically-protected surface states are expected, and we can unambiguously identify nontrivial surface states in the surface spectrum for the semi-infinite (010) surface, as displayed in Fig.~\ref{fig1}(h), confirming further the nontrivial topology in the AFM state.

\begin{table*}[tbp]
\centering
\caption{\label{tab1}Parameters derived from quantum oscillations for EuAs$_3$.}
\resizebox{\textwidth}{!}
{
\begin{tabular*}{17cm}{@{\extracolsep{\fill}}cccccccc}
\hline
\hline
& $F$ (T)& $E_F$ (meV)& $A_F$ (10$^{-3}$${\rm {\mathring {A}}}$$^{-2}$)& $k_F$ (10$^{-2}$${\rm {\mathring {A}}}$$^{-1}$)& $v_F$ (10$^5$ m/s)& $m^*$ ($m_0$)& $T_D$ (K)\\ 
\hline
$\alpha$& 156& 0.6& 14.9& 6.9& 1.3& 0.580(1)& 15.2(4)\\ 
$\beta$& 185& 0.8& 17.7& 7.5& 1.8& 0.45(1)& 5.0(6)\\ 
$\gamma^1$& 217& 1.0& 20.8& 8.1& 2.3& 0.38(3)& 8(1)\\ 
$\gamma^2$& 7& 2.1& 0.3& 1.0& 0.6& 0.178(6)& 8.0(5)\\ 
$\xi$& 93& 4.0& 8.9& 5.3& 1.7& 0.37(1)& 12(2)\\ 
$\alpha^{'}$& 158& 3.1& 15.2& 6.9& 1.6& 0.51(1)& 5.9(1)\\ 
$\varepsilon$& 346& 4.5& 33.3& 10.3& 3.6& 0.329(4)& 13.1(6)\\ 
$\eta$& 597& 4.0& 57.4& 13.5& 4.2& 0.370(3)& 9.8(7)\\ 
\hline
\hline
\end{tabular*}
} 
\end{table*}

\subsection {Electrical transport in the AFM state}

To verify the predictions from band calculations, we conducted electrical transport measurements. Resistivity in zero magnetic field is plotted in Fig.~\ref{fig2}(a), which displays typical metallic behaviour with a low-temperature peak corresponding to the magnetic transitions. The magnetic transitions were also found by thermodynamic measurements (Fig.\ S1) to be consistent with previous reports \cite{EuAs3_1,EuAs3_2,EuAs3_3,EuAs3_4,EuAs3_5,EuAs3_6}. The inset to Fig.~\ref{fig2}(a) shows the fit to the resistivity data below 2.5 K using a power law: $\rho$ = $\rho_0$+$AT^2$, where $\rho_0$ is the residual resistivity, and $A$ the electronic scattering coefficient. The fit gives a residual resistivity $\rho_0$ of 2.6\,$\mu\Omega$cm, and the residual resistivity ratio (RRR) $\rho_{300K}$/$\rho_0$ is $\sim$ 72. Figure~\ref{fig2}(b) shows the low-field MR data with evident SdH oscillations. $B_M$ in Fig.~\ref{fig2}(b) denotes the critical field, above which the spins are fully polarized by the external magnetic field. The SdH oscillation amplitude can be described by the Lifshitz-Kosevich (LK) formula \cite{Armitage,Burkov}
\begin{displaymath}
\Delta\rho_{xx}\propto\frac{2\pi^2k_BT/\hbar\omega_c}{\sinh(2\pi^2k_BT/\hbar\omega_c)}e^{-\frac{2\pi^2k_BT_D}{\hbar\omega_c}}\cos2\pi\left(\frac{F}{B}-\gamma +\delta\right)\text{,}
\end{displaymath}
where $\omega_c$ = $eB/{m^*}$ is the cyclotron frequency and $T_D$ is the Dingle temperature.
$\gamma$ = 1/2$-$(1/2$\pi$)$\phi_B$ (0 $\le$ $\gamma$ $\le$ 1) is the Onsager phase factor, and $\phi_B$ is a geometrical phase known as the Berry phase. For a topological system with peculiar electron state degeneracy and intraband coupling, a $\pi$ Berry phase will be observed. $2\pi\delta$ is an additional phase shift resulting from the curvature of the Fermi surface in the third dimension, where $\delta$ varies from 0 to $\pm 1/8$ for a quasi-2D cylindrical Fermi surface and a corrugated 3D Fermi surface, respectively \cite{SrAs3-1, SrAs3-2}. The cyclotron effective mass $m^*$ can be obtained from the thermal damping factor $R_T$ = $2\pi^2k_BT/\hbar\omega_c\sinh(2\pi^2k_BT/\hbar\omega_c)$.

By analyzing the oscillatory component (inset to Fig.~\ref{fig2}(c)) below $B_M$ via fast Fourier transform (FFT), four bands are uncovered, i.e., 156, 185, 217 and 7\,T, referred to as $\alpha$, $\beta$, $\gamma^1$, and $\gamma^2$, respectively, in line with the band calculations. To check their topological nature, a Landau index fan diagram is plotted in Fig.~\ref{fig2}(d), yielding intercepts of 0.6(2), 0.5(1), $-$0.03(9), and 0.07(8) for $\alpha$, $\beta$, $\gamma^1$ and $\gamma^2$, respectively. Throughout this paper, we assign integer indices to the $\Delta\rho_{xx}$ peak positions in 1/$B$ and half integer indices to the $\Delta\rho_{xx}$ valleys. According to the Lifshitz-Onsager quantization rule for a corrugated 3D Fermi surface, intercepts falling between $-$1/8 and 1/8 suggest nonzero Berry phase, while intercepts in the range 3/8--5/8 indicate trivial band topology. Therefore, the $\gamma^1$ and $\gamma^2$ bands are topologically protected, while the other two are trivial. Other parameters for these four bands, such as the Fermi energy $E_F$, extremal cross-sectional areas $A_F$, Fermi momentum $k_F$, Fermi velocity $v_F$, cyclotron effective mass $m^*$ and Dingle temperature $T_D$, are calculated and summarized in Table~\ref{tab1}.

In TSMs, in addition to nonzero Berry phase, the $n$-LMR induced by the chiral anomaly can also serve as a smoking gun for nontrivial topology \cite{nMR-1, nMR-2, nMR-3}. Figure~\ref{fig2}(e) displays the $n$-LMR of EuAs$_3$ with magnetic field parallel to the electric current $I$. The kinks in the $n$-LMR curves below the ordering temperature arise from the field-induced transitions \cite{EuAs3_5, EuAs3_6, EuAs3-MR}, i.e., from a collinear antiferromagnetic phase to incommensurate and commensurate spiral phases \cite{EuAs3_5, EuAs3_6, EuAs3-MR}. Negative MR in magnetic systems is not uncommon \cite{nMR-1, nMR-2, nMR-3}, when the applied magnetic field suppresses the inelastic magnetic scattering from local moments or magnetic impurities, leading to a negative MR for charge transport along all directions \cite{nMR-1, nMR-2, nMR-3}. However, in EuAs$_3$ we only observed a $n$-LMR when the magnetic field is applied parallel to the electric current (Fig.~S2). Furthermore, if the applied external magnetic field has a strong effect on the magnetic scattering and induces a $n$-LMR, the changes in the $n$-LMR will occur predominantly across the ordering temperature.  This is not observed.  Instead, we find several minor kinks arising from the magnetic transitions, on top of a much larger signal. Therefore, the suppression of magnetic scattering  can be excluded as the origin of the $n$-LMR in EuAs$_3$. The $n$-LMR also displays a wide variety of temperature dependences, ruling out current jetting effects and the weak localization effect \cite{nMR-1, nMR-2, nMR-3}. Since nontrivial band topology has been confirmed in EuAs$_3$, the chiral anomaly arising from Weyl fermions is the most likely mechanism behind the $n$-LMR.

\begin{figure*}
\includegraphics[clip,width=0.80\textwidth]{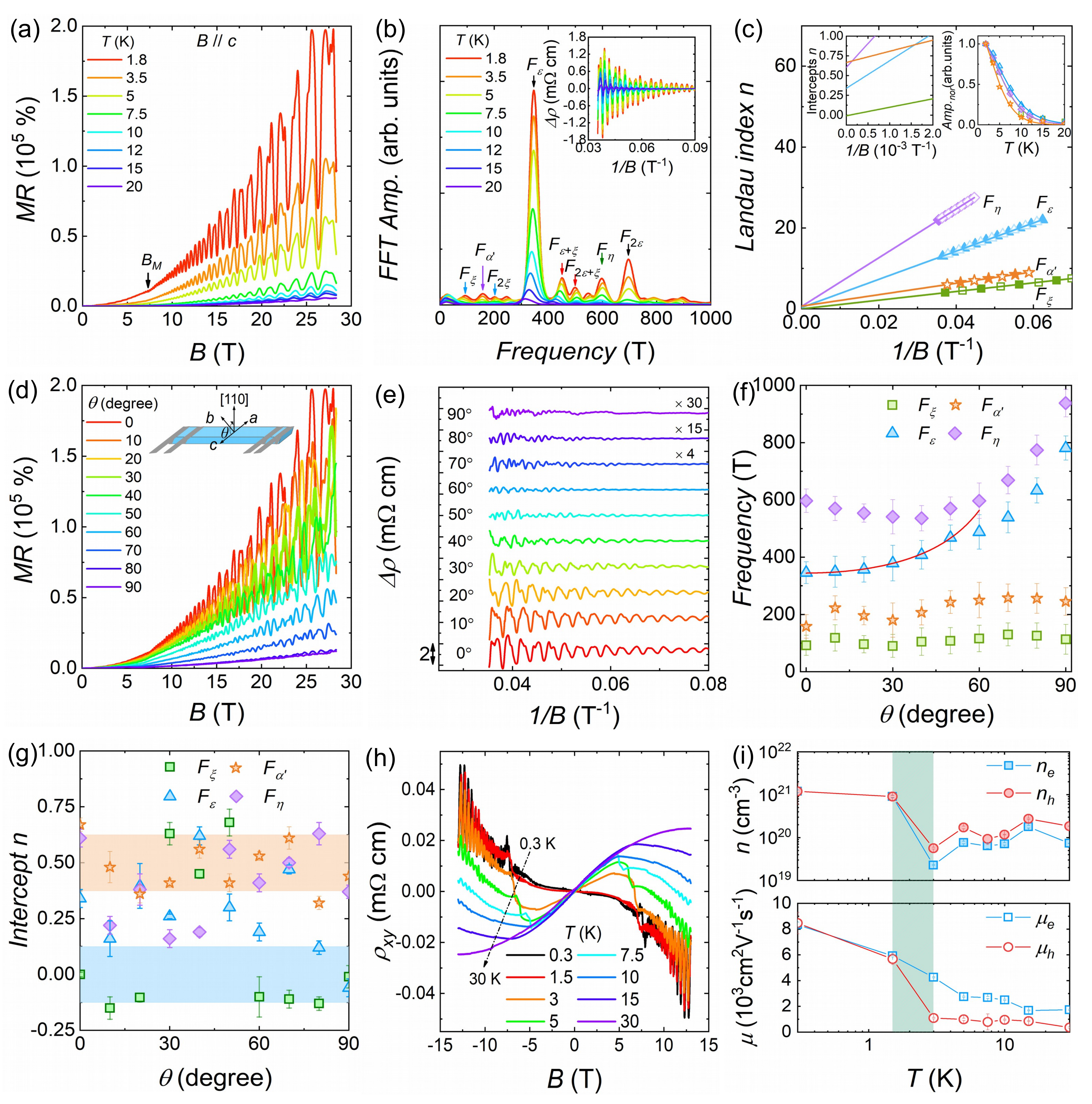}
\caption{\label{fig3}
Quantum oscillations study in the spin-polarized state of EuAs$_3$, and Hall resistivity measurements.
(a) Magnetoresistance measurements of EuAs$_3$ single crystal under higher magnetic field up to 28.3 T. 
(b) FFT results at various temperatures, yielding the four bands $\xi$, $\alpha'$, $\varepsilon$, and $\eta$. The inset displays the oscillatory component $\rho_{xx}$ above $B_M$. 
(c) Landau index $n$ plotted against 1/$B$ for the SdH oscillations at 1.8 K. The left inset shows the extrapolation of 1/$B$ to zero. The right inset shows the normalized FFT amplitude (Amp.$_{nor}$) as a function of temperature, and the solid lines represent the Lifshitz-Kosevich formula fit.
(d) SdH oscillations at different angles; the inset is a schematic illustration of the experimental geometry and the angle $\theta$. For $\theta$ = 0$^\circ$, the magnetic field is parallel to the $c$ axis. For $\theta$ = 90$^\circ$, the magnetic field is applied along the [110] direction.
(e) The oscillatory component $\rho_{xx}$ as a function of 1/$B$. The inset expands the data at 80 and 90$^\circ$.
Angular dependence of (f) the FFT frequencies, where error bars represent the full widths at half maximum of the FFT peaks, and (g) the Landau level index intercepts.
(h) Hall resistivity at various temperatures. 
(i) Carrier concentration and mobility as a function of temperature. The shadow area represents the temperature interval where a Lifshitz transition takes place. 
}
\end{figure*}

The $n$-LMR induced by the chiral anomaly in TSMs can be analyzed through the Adler-Bell-Jackiw (ABJ) chiral anomaly equation $\sigma(B) = (1+C_wB^2)(\sigma_0+a\sqrt{B})+\sigma_N$, where $\sigma_0$, $C_w$ and $\sigma_N = 1/(\rho_0+AB^2)$ denote the conductivity at zero field, a temperature-dependent positive parameter originating from the chiral anomaly, and the conventional nonlinear band contribution around the Fermi energy, respectively \cite{nMR-1, nMR-2, nMR-3}. Figure~\ref{fig2}(f) shows the conductivity and the fit to the data below 3 T at various temperatures. The data above the ordering temperature are well described by the ABJ equation, while the data at lower temperatures don't fit as well, which may be ascribed to the effect that magnetic transitions have on the chiral current. The inset to Fig.~\ref{fig2}(f) shows the temperature dependence of $C_w$. At 2 K, $C_w$ is 0.253(7) T$^{-2}$. With increasing temperature, a clear anomaly in $C_w$ around the ordering temperature can be observed, verifying the proposal above. When $T$ $>$ $T_N$, $C_w$ decreases monotonically, as observed in SrAs$_3$ \cite{SrAs3-1}. Taken together, these results demonstrate that EuAs$_3$ is a magnetic topological massive Dirac metal in its AFM ground state.

\begin{figure*}
\includegraphics[clip,width=\textwidth]{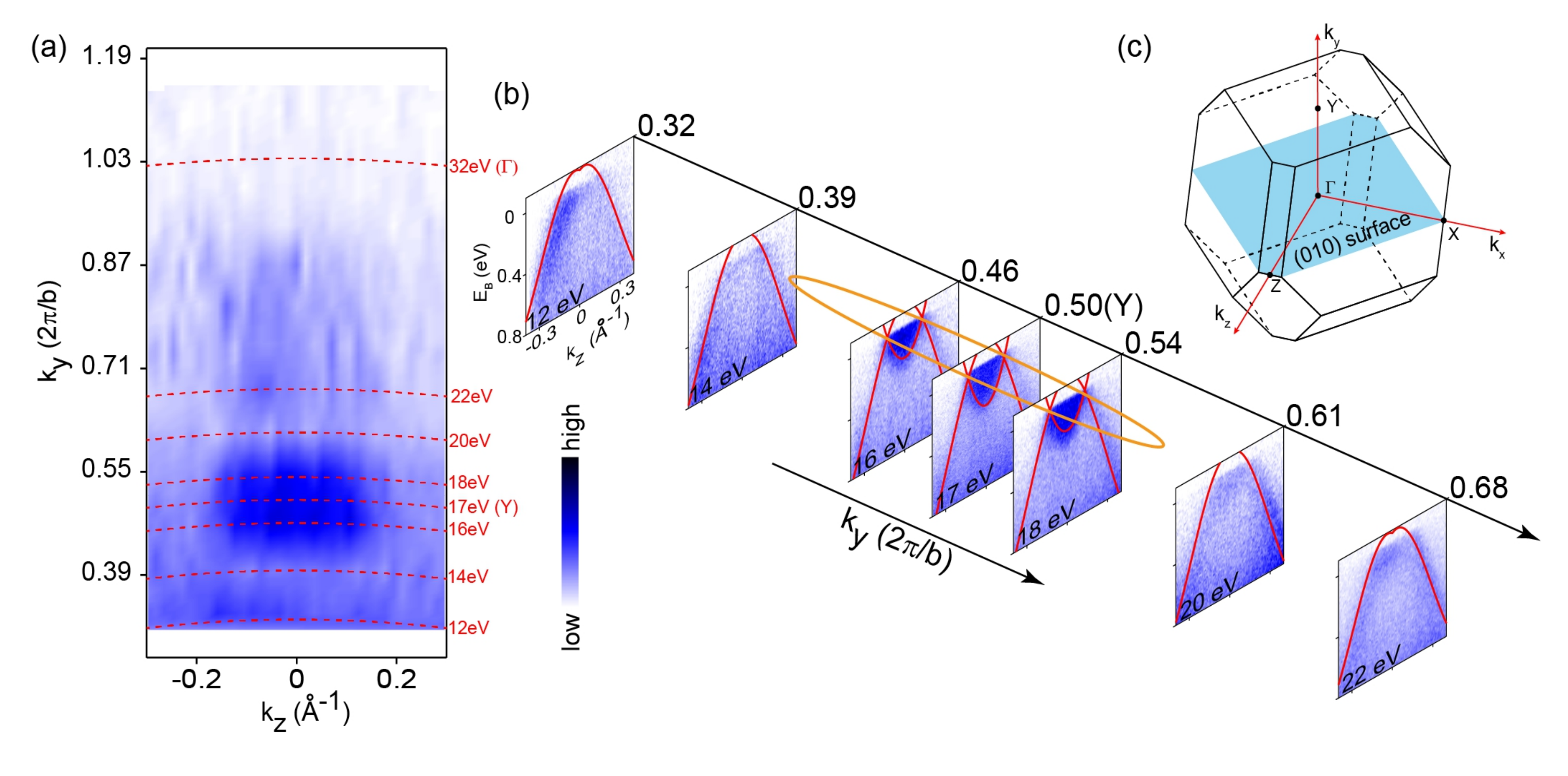}
\caption{\label{fig4}
Verification of topological nodal-line structure by ARPES measurements in the paramagnetic state of EuAs$_3$.
(a) Photon-energy-dependent plot of photoemission intensities at the Fermi surface along the $k_z$ direction. The red dotted lines denote the $k_y$ momentum locations probed by different photon energies.
(b) The dispersions along the $k_z$ direction probed by different photon energies corresponding to different $k_y$. The calculated electronic structure is superimposed as red curves. The orange ellipse illustrates the topologically nontrivial nodal loop.
(c) The Brillouin zone of EuAs$_3$, with high-symmetry points and (010) surface labeled.
}
\end{figure*}

\section{Topological nodal-line structure in the spin-polarized and paramagnetic states}

\subsection {Electrical transport measurements in the spin-polarized state}

We now turn to the exploration of topology in the spin-polarized state, where in Fig.~\ref{fig1}(b) we have already observed a clear change in the quantum oscillations. Figure~\ref{fig3}(a) plots the MR of EuAs$_3$ in higher magnetic field, and an unstaturated XMR $\sim$ 2$\times$10$^5$ $\%$ at 1.8 K and 28.3 T is observed. By analyzing the oscillatory components above $B_M$ (inset in Fig.~\ref{fig3}(b)), frequency components are identified at $F$ = 93, 158, 346, and 597 T, which are referred to here as the $\xi$, $\alpha^{'}$, $\varepsilon$, and $\eta$ bands, respectively (Fig.~\ref{fig3}(b)). These four bands are different from those in lower field (Fig.~\ref{fig2}(c)), indicating that they are likely rooted in different band structure. This is unsurprising since the unit cell is no longer doubled by antiferromagnetism, but the field-induced spin polarization can also play a significant role.  We thus conducted band calculations for the field-polarized state (Fig.~S3) and the paramagnetic state (Fig.~S4), and these are indeed quite different, as we discuss in more detail in the Supplementary Information. In the field-polarized state we find four Fermi surface sheets---two electron and two hole sheets---and double nodal loops at the $Y$ point, one each for the spin-up and spin-down channels (Fig.~S3(d)). To identify the topological nature of the four bands seen in quantum oscillations, a Landau index fan diagram is plotted in Fig.~\ref{fig3}(c), and the intercepts are -0.0(1), 0.67(3), 0.34(4), and 0.61(4) for the $\xi$, $\alpha^{'}$, $\varepsilon$, and $\eta$ bands, respectively. The intercepts indicate that the $\xi$ band is topologically protected, while the $\alpha^{'}$ and $\eta$ bands are topologically trivial. The intermediate value for $\varepsilon$ is suggestive of a nontrivial Berry phase, but does not allow a strong conclusion, and will require further verification. The cyclotron effective masses $m^*$ for these four pockets can be obtained by fitting the temperature dependence of the normalized FFT amplitude, as shown in the right inset to Fig.~\ref{fig3}(c). Other parameters can be also extracted, and all values are summarized in Table~\ref{tab1}.

To better reveal the Fermi surface anisotropy and topology of EuAs$_3$, angle-dependent MR measurements have been performed at 1.8 K, in the experimental geometry shown in the inset to Fig.~\ref{fig3}(d). Upon rotating sample from 0$^\circ$ to 90$^\circ$, the magnitude of the MR is reduced monotonically, as also seen in a polar plot of the MR (Fig.~S2(b)). We extract the frequency components for the $\xi$, $\alpha^{'}$, $\varepsilon$, and $\eta$ bands by analyzing the oscillatory component (Fig. 3(e)) and summarize the results in Fig.~\ref{fig3}(f). The angle dependence of the $\xi$, $\alpha^{'}$, and $\eta$ bands is of 3D character, while the $\varepsilon$ band is well described below 50$^\circ$ by the formula $F$ = $F_{3D}$+$F_{2D}$/cos($\theta$), where $F_{2D}$ and $F_{3D}$ denote 2D and 3D components. The ratio between the 2D and 3D components derived from the fit ($F_{2D}$/$F_{3D}$) is $\sim$ 1.76, suggesting that the $\varepsilon$ pocket exhibits mainly 2D character although a 3D component also exists.

Now, we turn to the angle dependence of the Berry phase. As shown in Fig.~\ref{fig3}(g), the intercept for the topological $\xi$ band shows strong angle dependence, similar to results in other systems such as Cd$_3$As$_2$ \cite{Cd3As2-angle}, ZrSi$M$ ($M$ = Se, Te) \cite{ZrSiS}, or ZrTe$_5$ \cite{ZrTe5}. For $\theta$ $<$ 30$^\circ$ and $\theta$ $\ge$ 60$^\circ$, the intercept falls between $-$1/8 and 1/8, while it falls between 3/8 and 5/8 for 30$^\circ$ $\le$ $\theta$ $<$ 60$^\circ$. For the $\varepsilon$ band, the intercept from 0$^\circ$ to 70$^\circ$ fluctuates between 1/8 and 5/8, averaging to 0.34(5), which is suggestive of trivial topology. However, when $\theta$ reaches 80$^\circ$ and 90$^\circ$, this intercept falls between $-$1/8 and 1/8, implying nonzero Berry phase. For $\eta$ and $\alpha^{'}$ bands, the intercepts at all angles remain between 1/8 and 5/8, averaging 0.4(2) and 0.5(2), respectively, indicating trivial topology.

\subsection {ARPES measurements in the paramagnetic state}

Since our band structure calculations identify nodal loops at $Y$ and our quantum oscillation data indicate nontrivial band topology, we also directly investigated the band structure with ARPES.  Momentum analysis in this technique is incompatible with magnetic field, so we investigated the paramagnetic rather than the field-polarized state; however, as shown in more detail in supplementary Figs.~S3 and S4, a closed nodal loop persists at the $Y$ point in the paramagnetic state. In order to visualize the nodal loop in EuAs$_3$, the photon energy dependence of the electronic structure along the $k_y$ direction was investigated at 18\,K within the vertical plane of the (010) cleaved surface, as sketched in Fig.~\ref{fig4}(c). From the intensity plot of the Fermi surface in the $k_y$--$k_z$ plane (Fig.~\ref{fig4}(a)), the pocket centered at the $Y$ point (17\,eV) can be easily identified, and two nodes arising from the crossing of the electronlike and holelike bands can be also observed in Fig.~\ref{fig4}(b), which agrees with the band calculations (red lines). We estimate the Fermi momenta $k_F$ and Fermi velocities $v_F$ to be $k_F$ = 0.12 and 0.14\,\AA$^{-1}$ and $v_F$ = 3.7$\times 10^5$ and 1.17$\times 10^5$ m/s, respectively, for the hole and electron bands, the same order of magnitude as for the $\varepsilon$ and $\eta$ bands (see Table~\ref{tab1}). For ARPES cuts away from the $Y$ point (Fig.~\ref{fig4}(b)), the band-crossing area shrinks gradually and finally disappears. This is highlighted by an orange ellipse, and represents the nodal loop predicted by our band structure calculations. These data are extremely similar to what was found in SrAs$_3$ \cite{SrAs3-3}.

The verification of the nodal-line structure in the paramagnetic state serves as strong evidence for the existence of nodal-line structure in the spin-polarized state, which has closely similar but spin-split band structure, but the similarities may not end there. Very recently, lifted degeneracy of the Bloch bands was observed in the paramagnetic phase of EuCd$_2$As$_2$, producing a spin-fluctuation-induced Weyl semimetal state \cite{EuCd2As2}. The magnetic susceptibility in EuAs$_3$ reveals a positive Curie-Weiss temperature $T_{CW}$ of 4.4\,K for magnetic fields applied within the $ab$ plane (Supplementary Fig.~S6), suggestive of ferromagnetic fluctuations deep in the paramagnetic phase. The ferromagnetic correlations in EuAs$_3$ may induce band splitting within the paramagnetic phase, which may be resolvable with higher-resolution ARPES, such as laser ARPES.

\section{discussion and conclusion}

It is clear from our transport measurements that the electronic structure in the antiferromagnetically ordered state is very different from that found in the field-polarized or paramagnetic states.  This is a consequence of the doubling of the unit cell due to antiferromagnetic order and the coupling of this magnetic order to the electronic structure, and is well described by our band calculations.  However, we also find evidence for an additional Lifshitz transition within the antiferromagnetic phase.

Figure~\ref{fig3}(h) shows the Hall resistivity ($\rho_{xy}$) from 0.3 to 30 K. The $\rho_{xy}$ curves are clearly nonlinear, implying the coexistence of two types of carriers, as predicted by band structure calculations. On cooling, the slope of the curve changes from positive to negative, indicating an increased contribution from electron carriers. The carrier concentration and mobility are extracted by fitting the Hall conductivity with a two-carrier model \cite{SrAs3-2}, and results are summarized in Fig.~\ref{fig3}(i). For 3 $\le T\le$ 30 K, the concentration of hole carriers is larger than that of electron carriers. Upon decreasing the temperature below 3 K, the concentration of electron carriers is suddenly enhanced, accompanied by a sharp increase in the mobility of hole carriers. These indicate a possible Lifshitz transition. To check this, we analyze the oscillatory component ($\Delta\rho_{xy}$), and identify a new oscillation frequency of 374 T (denoted as the $\phi$ band) with trivial topology (see Fig.~S5(b)), demonstrating that a Lifshitz transition does indeed occur. Temperature-induced Lifshitz transitions are also observed in other TSMs, for example, $M$Te$_5$ ($M$=Zr, Hf) \cite{ZrTe5-LT,HfTe5-LT}, which have been used to explain the origin of the resistivity anomaly. However, no such anomaly can be observed in EuAs$_3$, indicative of its unusual origin. Generally speaking, Lifshitz transitions are related to electronic transitions at zero temperature, and involve abrupt changes of the Fermi surface topology. However, in topological materials, Lifshitz transitions can also involve other types of zero-energy modes, such as Weyl or Dirac nodes, nodal lines, flat bands, Majorana modes, etc.~\cite{LT-1}. It has been proposed that multiple types of novel Lifshitz transitions involving Weyl points are possible depending on how they connect Fermi surfaces and pockets. For instance, the Lifshitz transition can correspond to the transfer of Berry flux between Fermi pockets connected by type-II Weyl points \cite{LT-2}. To understand the physics behind the apparent low-temperature Lifshitz transition in EuAs$_3$, more work is needed.

According to the conventional charge-carrier compensation picture for XMR, the ratio $n_h/n_e$ should be unity \cite{Armitage, Burkov}. At 0.3 K, $n_h/n_e$ for EuAs$_3$ is $\sim$ 1.0, consistent with this picture. However, for 3 $\le T \le$ 30 K, $n_h/n_e$ varies between 1.5 and 2.5 while the MR remains large and unsaturated, evidently excluding the charge-compensation picture for EuAs$_3$. XMR is also frequently encountered in the cases of topologically-protected electronic band structure, and when open orbits contribute \cite{Armitage,Burkov,XMR-1,XMR-2}. According to the open-orbit effect, the unsaturated XMR is only observed for current along the open orbits \cite{XMR-2}. However, the unsaturated XMR with different current direction in EuAs$_3$ excludes the open-orbit effect (see Fig. \ S7). Beside, for both the charge-carrier compensation picture and open-orbit effect, a $B^2$ dependence of MR is suggested \cite{XMR-1, XMR-2}, which is different from the situation of EuAs$_3$ reported here (Fig. \ S7(c)). Since we have verified nontrivial band topology in EuAs$_3$, we consider this the more likely explanation.

In summary, combining band calculations, electrical transport and ARPES measurements on the magnetic compound EuAs$_3$, we report a magnetism-induced topological transition from a TNL semimetal in the paramagnetic or the spin-polarized state to a topological massive Dirac metal in the AFM ground state. The paramagnetic and spin-polarized states differ by the splitting of a topological nodal line associated with the spin-splitting of the band structure.  An XMR of $\sim 2\times 10^5$\,\% and an as-yet-unexplained temperature-induced Lifshitz transition below 3\,K have also been revealed. These results indicate that magnetic EuAs$_3$ could serve as a unique platform to explore exotic physics at the interface of magnetism and topology.

\begin{acknowledgments}
This work is supported by the Ministry of Science and Technology of China (Grant Nos.\ 2015CB921401 and 2016YFA0300503), the Natural Science Foundation of China (Grant Nos.\ 11421404, 11674367, 11674229 and 11874264), the NSAF (Grant No.\ U1630248), and the Zhejiang Provincial Natural Science Foundation (Grant No.\ LZ18A040002), the National Key R$\&$D program of China (Grant No. 2017YFA0305400) and the Ningbo Science and Technology Bureau (Grant No.\ 2018B10060). W.W.Z.\ is supported by the Shenzhen Peacock Team Plan (KQTD20170809110344233) and Bureau of Industry and Information Technology of Shenzhen through the Graphene Manufacturing Innovation Center (201901161514). Y.F.G.\ acknowledges research funds from the State Key Laboratory of Surface Physics and Department of Physics, Fudan University (Grant No.\ KF2019\_06). C.Y.X.\ was supported by the Users with Excellence Project of Hefei Science Center CAS (Grant No.\ 2018HSC-UE015). D.C.P.\ is supported by the Chinese Academy of Sciences through 2018PM0036 and from the Deutsche Forschungsgemeinschaft (DFG), through project C03 of the Collaborative Research Center SFB 1143 (project-ID 247310070). The authors are grateful for support from the Analytical Instrumentation Center (\#\ SPST-AIC10112914), SPST, ShanghaiTech University.
\end{acknowledgments}

\appendix
\renewcommand\thesection{\Roman{section}}
\renewcommand\thesubsection{\Alph{subsection}}
\section*{Methods and Extended Data}

\section{Methods}

\subsection{Sample synthesis}
Eu (99.95\%, Alfa Aesar), As (99.999\%, PrMat) and Bi (99.9999\%, Aladdin) blocks were mixed in a molar ratio of 1:3:26 and placed into an alumina crucible. The crucible was sealed in a quartz ampoule under vacuum and subsequently heated to 900\,$^\circ$C in 15\,h. After reaction at this temperature for 20\,h, the ampoule was cooled to 700\,$^\circ$C over 20 h, and then slowly cooled to 450\,$^\circ$C at $-$1\,$^\circ$C/h. The excess Bi flux was then removed using a centrifuge, revealing EuAs$_3$ single crystals with black shiny metallic lustre.

\subsection{Electrical transport and thermodynamic measurements}

For electrical transport measurements, a single crystal was cut into a bar shape with current applied in the $ab$ plane. A standard four-probe method was used for the longitudinal resistivity measurement. Data were collected in a $^3$He and a $^4$He cryostat. Magnetic susceptibility and specific heat measurements were performed in a magnetic property measurement system (MPMS, Quantum Design) and a physical property measurement system (PPMS, Quantum Design), respectively. High-field measurements were performed at the Steady High Magnetic Field Facilities, High Magnetic Field Laboratory, Chinese Academy of Sciences, in Hefei.

\subsection{Angle-resolved photoemission spectroscopy (ARPES) measurements}

ARPES measurements were performed at beamline BL13U at the National Synchrotron Radiation Laboratory (NSRL), China (photon energy $h\nu$ = 12-38\,eV). The samples were cleaved in situ and measured in ultrahigh vacuum with a base pressure of better than 3.5 $\times$ 10$^{-11}$ mbar, and data were recorded by a Scienta R4000 analyzer with the sample at 18 K. The energy and momentum resolution were 10\,meV and 0.2$^\circ$.

\subsection{Density functional theory (DFT) calculations}

First-principles calculations were carried out within the framework of the projector augmented wave (PAW) method \cite{18,19}, and employed the generalized gradient approximation (GGA) \cite{20} with Perdew-Burke-Ernzerhof (PBE) formula \cite{21}, as implemented in the Vienna {\itshape ab initio} Simulation Package ({\sc Vasp}) \cite{22}. Two unit cells repeated along the $b$ axis were adopted to simulate the antiferromagnetic configuration indicated by neutron diffraction experiment for EuAs$_3$ \cite{EuAs3_2}. The energy cutoff was chosen to be 500 eV. A $\Gamma$-centered $8\times6\times14$ Monkhorst-Pack $k$-point grid was used to produce the well-converged results for the antiferromagnetic phase. For the spin-polarized and paramagnetic band calculations, the same unit cell was used. $\Gamma$-centered $10\times10\times10$ and $6\times6\times6$ grids were used in the first Brillouin zone for the unit cell and supercell magnetic structures, respectively. The convergence criterion of energy in relaxation was set to be 10$^{-6}$\,eV and the atomic positions were fully relaxed until the maximum force on each atom was less than 0.02\,eV/\AA. The electronic correlations of Eu-4$f$ states were treated by the GGA + $U$ method \cite{23}. SOC was considered in a self-consistent manner. The {\sc Wannier90} package \cite{24} was adopted to construct Wannier functions from the first-principles results. The {\sc WannierTools} code \cite{25} was used to investigate the topological features of surface state spectra.

\section{Extended Data}

\subsection{X-ray diffraction and thermodynamic measurements}

Figure S1(a) shows the powder x-ray diffraction (XRD) pattern from the largest surface of a EuAs$_3$ single crystal. The inset shows the x-ray rocking curve of the (002) peak, the full width at half maximum (FWHM) of which is 0.06$^\circ$, indicative of the high quality of our single crystals. Figure S1(b) shows the verification of the magnetic transitions by resistivity, specific heat and magnetic susceptibility, evidencing the antiferromagnetic ($T_N$) and lock-in ($T_L$) transitions at 11 and 10.3 K, respectively, consistent with previous reports \cite{EuAs3_2}. Figure S1(c) displays specific heat measurements under magnetic fields. With increasing field, both $T_N$ and $T_L$ decrease. By extrapolating the data to zero temperature, as plotted in the inset to Fig.\ S1(c), $T_N$ and $T_L$ will disappear at $\sim$ 11.1 and $\sim$ 10.7\,T, respectively. To avoid any influence from the metamagnetic transition and investigate the fully spin-polarized state, we collected our quantum oscillation data above 11.1\,T (Figs.\ 3(b) and 3(e) in the main text).

\subsection{Angular-resolved magnetoresistance (AMR) in low magnetic field}

Figure S2(a) illustrates the experimental geometry for our AMR measurements. For $\theta$ = 0$^\circ$ and 90$^\circ$, the magnetic field is parallel to the $c$ axis and the [110] direction, respectively. For $\varphi$ = 0$^\circ$, the magnetic field is parallel to the [110] direction in the $ab$ plane, i.e., perpendicular to the electric current $I$. For $\varphi$ = 90$^\circ$, the magnetic field is parallel to the electric current $I$. Polar plots of the AMR for $\theta$ and $\varphi$ at several magnetic fields are displayed in Figs.\ S2(b) and S2(c), respectively, showing typical twofold anisotropy. Compared with $\theta$, the AMR for $\varphi$ is much more complicated, which is ascribed to the field-induced magnetic transitions \cite{EuAs3-MR}. For $\varphi \sim 90^\circ$, i.e., magnetic field parallel to electric current $I$, a negative MR is observed. According to Figs.\ S2(b) and S2(c), the scenario that external magnetic field suppresses the inelastic magnetic scattering from local moments or magnetic impurities and then induces a negative MR along all directions can be excluded.

\subsection{Band structure in the spin-polarized state}

The GGA+$U$ ($U$ = 5\,eV) band structures of the spin-up and spin-down electrons, shown in Figs.\ 3(a) and 3(b), exhibit topological semimetal states with band crossings near the Fermi level. For both spin-up and spin-down channels, the band crossings persist along a closed path around the $Y$ point in the BZ, showing a double nodal-line structure, as sketched in Fig.\ S3(d). When spin-orbit coupling (SOC) is considered, the band crossings will be fully gapped because of the band hybridization, as shown in Fig.\ S3(c). Figure 3(e) shows the Fermi surfaces of the spin-polarized EuAs$_3$ in presence of SOC. There are four main Fermi surface sheets---two hole sheets and two electron sheets. All these Fermi surfaces clearly display three-dimensional (3D) character.

\subsection{Band structure in the paramagnetic state}

In order to simulate the paramagnetic phase of EuAs$_3$, we treated the 4$f$ electrons on Eu as core electrons. In the absence of SOC, the overall band profiles near the Fermi level, shown in Fig.\ S4(a), share many key characteristics with those of SrAs$_3$ \cite{CaP3,SrAs3-1}, making EuAs$_3$ a topological nodal-line semimetal (TNLS) in the paramagnetic state. Our band structure calculations indicate a nodal loop around the $Y$ point in the BZ. When SOC is taken into consideration, gaps will be opened along the nodal line, and EuAs$_3$ becomes a small-gap insulator defined on curved Fermi levels. The calculated surface states in Fig.\ S4(d) illustrate a drumhead-like surface band, confirming further the TNLS character of paramagnetic EuAs$_3$. The Fermi surface in the paramagnetic state is calculated and displayed in Fig.\ S4(c), and consists of two large sheets around the $\Gamma$ and $T$ points, and one small sheet around the $Y$ point. The Fermi surface in the paramagnetic phase of EuAs$_3$ is very similar in shape to that of SrAs$_3$ \cite{SrAs3-1}, indicating that the nodal-line structure is quite robust, so direct verification of the nodal-line structure in the paramagnetic state by ARPES would serve as strong evidence for the existence of the nodal-line structure in the spin-polarized state.

\subsection{Quantum oscillations in the Hall resistivity}

To check the temperature-induced Lifshitz transition, we analyze the oscillatory component ($\Delta$$\rho_{xy}$) (displayed in the inset to Fig. S5(a)) via FFT, and a new band (denoted as $\phi$) with oscillation frequency of 374 T has been identified, demonstrating the Lifshitz transition. We then check the topology of the $\phi$ band, as plotted in Fig. S5(b). We assign integer indices to the $\Delta$$\rho_{xy}$ peak positions in 1/$B_{n+1/4}$ and half integer indices to the $\Delta$$\rho_{xy}$ valleys, where $n$+1/4 means the 1/4 phase shift in Hall resistivity compared with longitudinal resistivity \cite{LPHe}. The intercept of 0.65(2) falls in the range between 3/8 and 5/8, as shown in the inset to Fig. 4(d), suggesting the possibly trivial topology of the $\phi$ band.

\subsection{Susceptibility measurements}

To check whether ferromagnetic correlations exist in the paramagnetic state, the Curie-Weiss temperature $T_{CW}$ is extracted by fitting the temperature dependence of the inverse susceptibility to the Curie-Weiss law, as shown in Fig.\ S6. A positive value of 4.4\,K is extracted for magnetic fields applied in the $ab$ plane, evidencing possible ferromagnetic fluctuations.

\subsection{Exclusion of the open-orbit effect for XMR}

To exclude the open-orbit effect which may also induce the XMR in EuAs$_3$, we performed the MR measurements with electric current $I$ applied along different crystallographic axis, as shown in Fig.\ S7. Figure \ S7(a) shows the MR of Sample 2 with magnetic field parallel to $c$ axis and electric current applied along the [110] direction, which is the same as Sample 1 (Fig. \ 2(b)). Compared with Sample 1, the MR of Sample 2 with RRR $\sim$ 61 is smaller than that of Sample 1, indicating that the quality of single crystal may have a great effect on the magnitude of MR. For Sample 3 with smaller RRR $\sim$ 52, the MR is smaller than those of Sample 1 and Sample 2. For comparison, we plot the $B^2$ dependence of MR for these three samples, as shown in Fig. \ S7(c), which dispalys similar behavior in spite of the magnitude of MR (The difference may come from the different quality of single crystals.). According to the open-orbit effect, the unsaturated XMR is only observed for current along the open orbits, which means that changing the current direction will lead to a saturated MR \cite{XMR-2}. Therefore, the unsaturated XMR with different current direction in EuAs$_3$ suggests that the XMR resulting from the open-orbit effect can be excluded.

\begin{figure*}[h!]
\renewcommand\thefigure{S1}
\includegraphics[clip,width=\textwidth]{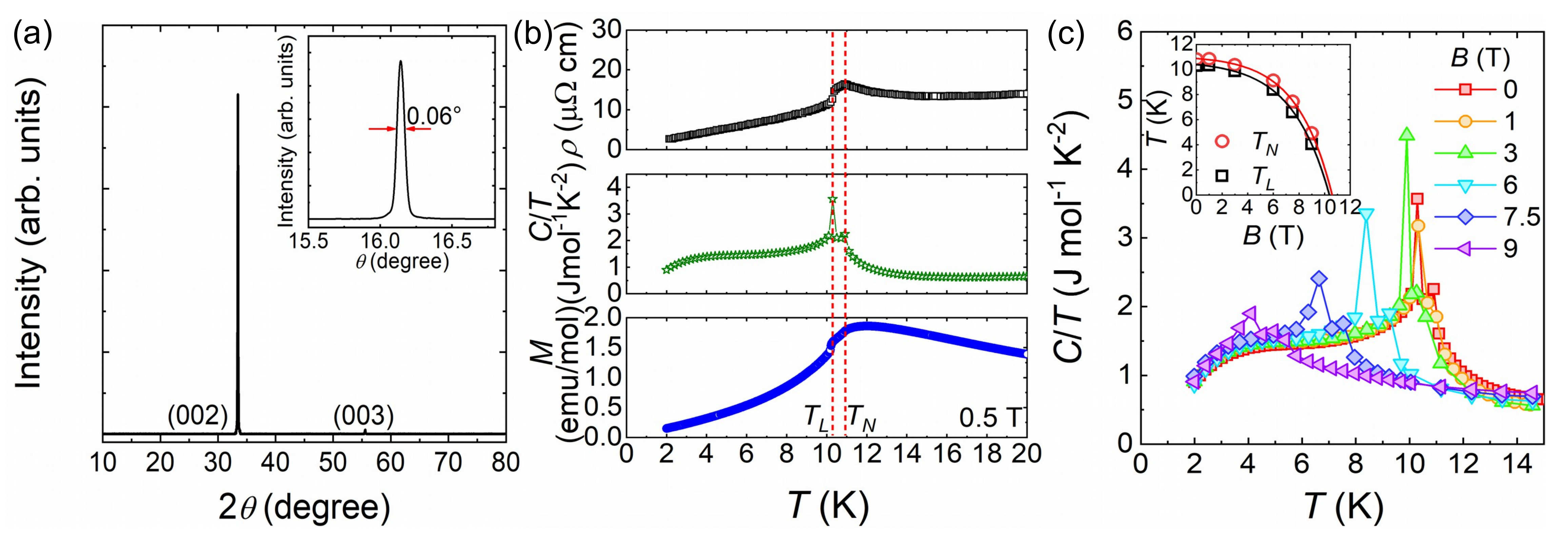}
\caption{\label{figS1}
Basic physical properties of the as-grown EuAs$_3$ single crystals.
(a) X-ray diffraction pattern from the largest surface of a EuAs$_3$ single crystal. The inset shows the rocking curve of the (002) peak.
(b) Antiferromagnetic ($T_N$) and lock-in ($T_L$) transitions demonstrated by resistivity, specific heat and magnetic susceptibility. 
(c) Specific heat measurements under magnetic fields applied along the $c$ axis.
}
\end{figure*}

\begin{figure*}[h!]
\renewcommand\thefigure{S2}
\includegraphics[clip,width=0.95\textwidth]{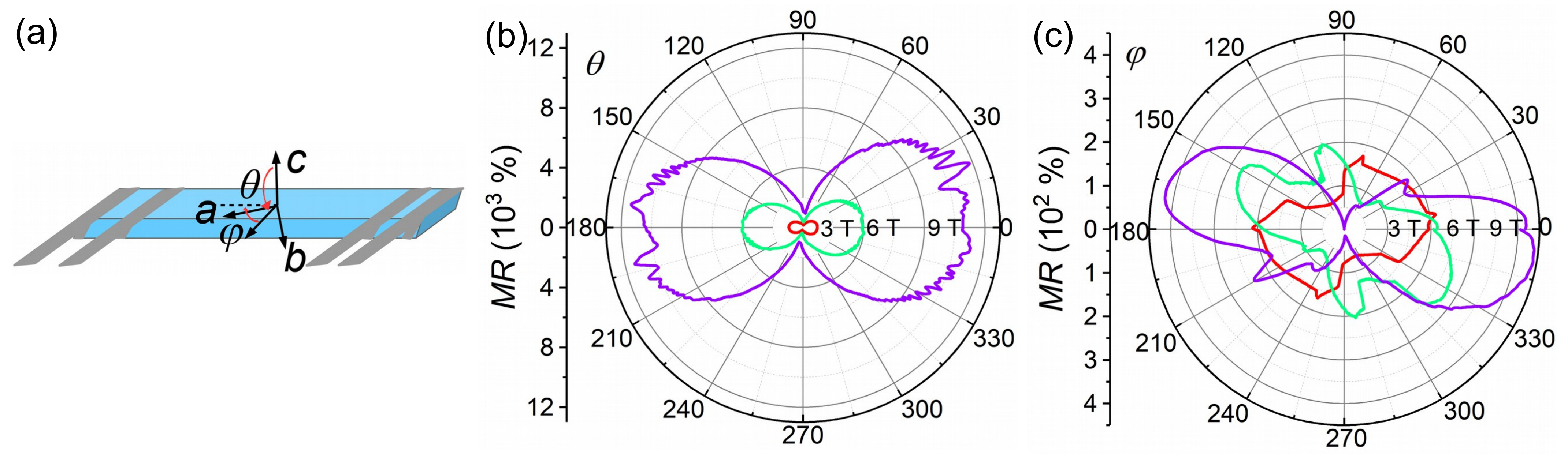}
\caption{\label{figS2}
Angular-resolved magnetoresistance (AMR) of EuAs$_3$ single crystal at 2 K.
(a) Schematic illustration of the experimental geometry and the angles $\theta$ and $\varphi$, and polar plots of the AMR for (b) $\theta$ and (c) $\varphi$ at several magnetic fields.
}
\end{figure*}

\begin{figure*}[h!]
\renewcommand\thefigure{S3}
\includegraphics[clip,width=0.9\textwidth]{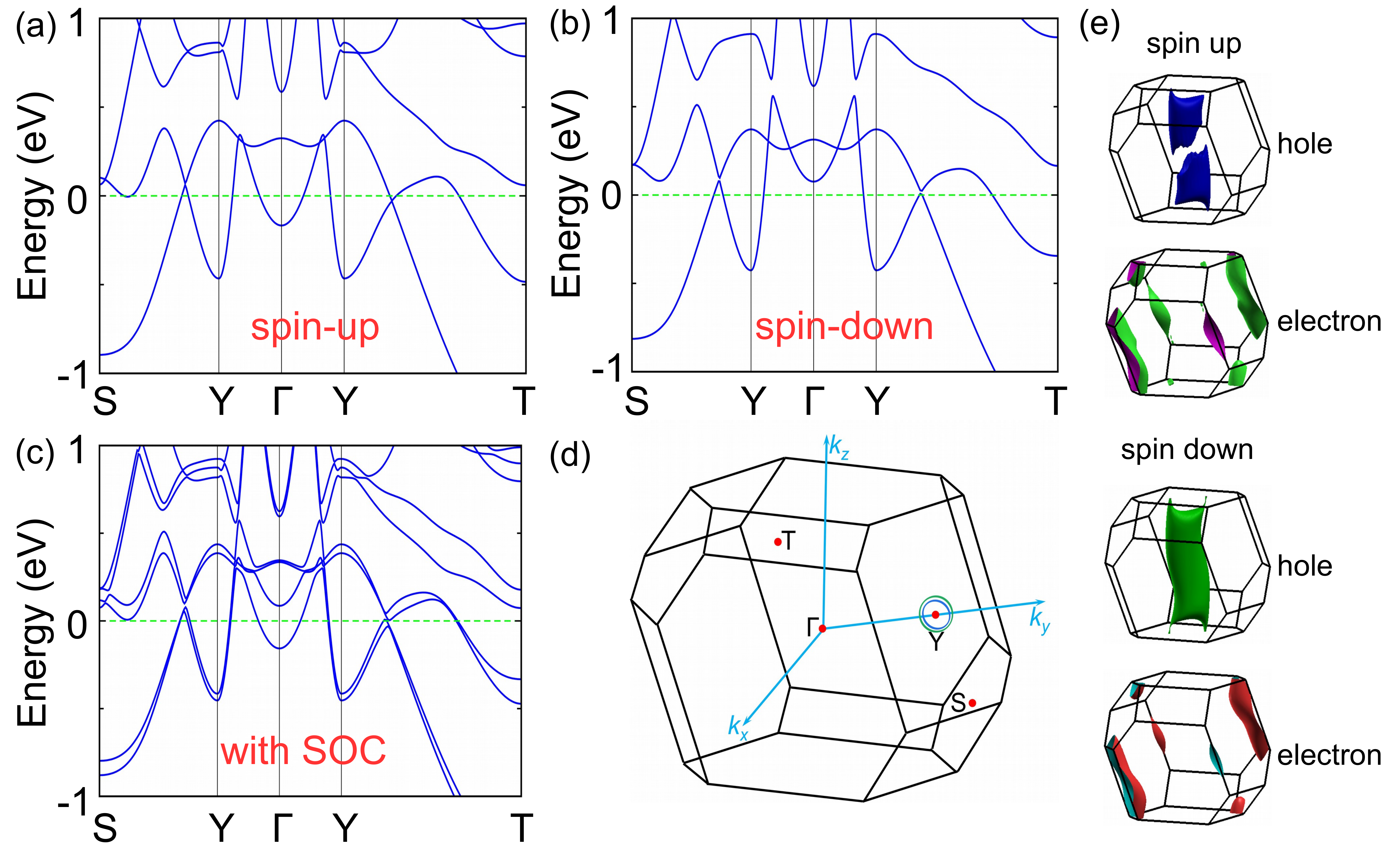}
\caption{\label{figS3}
Topological nodal-line structure in the spin-polarized state of EuAs$_3$. Band calculations for the spin-polarized state of EuAs$_3$: (a) spin-up, (b) spin-down and (c) spin-orbit coupling (SOC)-involved band structures. (d) The bulk Brillouin zone of spin-polarized EuAs$_3$. The green and blue loops surrounding the $Y$ point represent the nodal-line structure for the spin-up and spin-down channels, respectively. (e) Fermi surfaces of spin-polarized EuAs$_3$.
}
\end{figure*}

\begin{figure*}[h!]
\renewcommand\thefigure{S4}
\includegraphics[clip,width=0.68\textwidth]{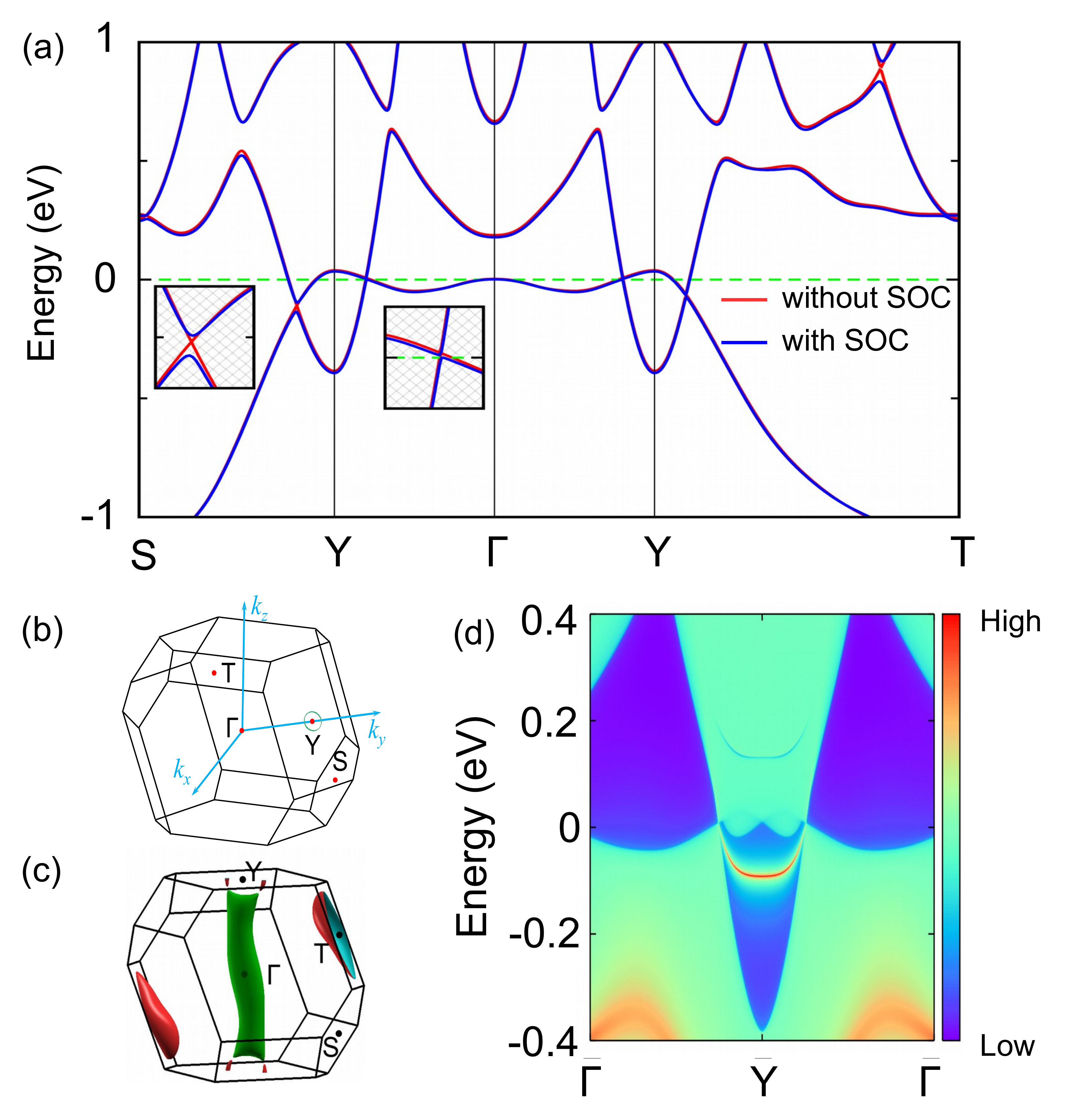}
\caption{\label{figS4}
Topological nodal-line structure in the paramagnetic state of EuAs$_3$. 
(a) Band structure of EuAs$_3$. 
(b) The bulk Brillouin zone of the paramagnetic EuAs$_3$. A single nodal loop surrounding the $Y$ point can be seen.
(c) Fermi surface of paramagnetic EuAs$_3$.
(d) The surface states of paramagnetic EuAs$_3$ on the projected surface perpendicular to the $k_z$ axis.
}
\end{figure*}

\begin{figure*}[h!]
\renewcommand\thefigure{S5}
\includegraphics[clip,width=0.7\textwidth]{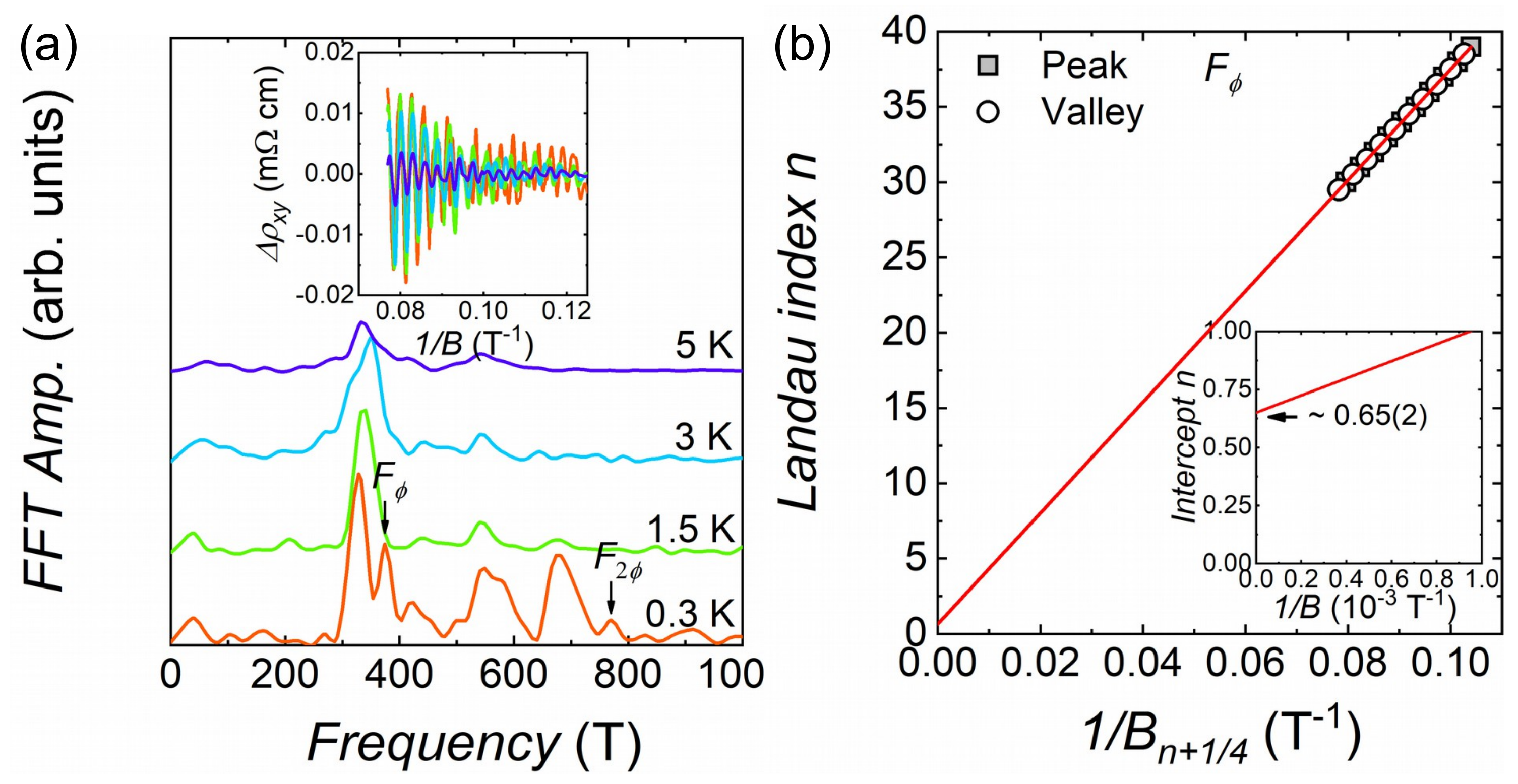}
\caption{\label{figS5}
The analysis of Hall resistivity oscillations of EuAs$_3$ single crystal.
(a) FFT results based on Hall resistivity oscillations at several temperatures. The inset shows the oscillatory component of $\rho_{xy}$ as a function of 1/$B$. A new frequency denoted as $F_{\phi}$ identifies a new $\phi$ band.
(b) Landau index $n$ as a function of 1/$B_{n+\frac14}$ in consideration of a 1/4 phase shift \cite {LPHe}. The red line represents a linear fit. The inset shows the extrapolation of 1/$B$ to zero.}
\end{figure*}

\begin{figure*}[h!]
\renewcommand\thefigure{S6}
\includegraphics[clip,width=0.7\textwidth]{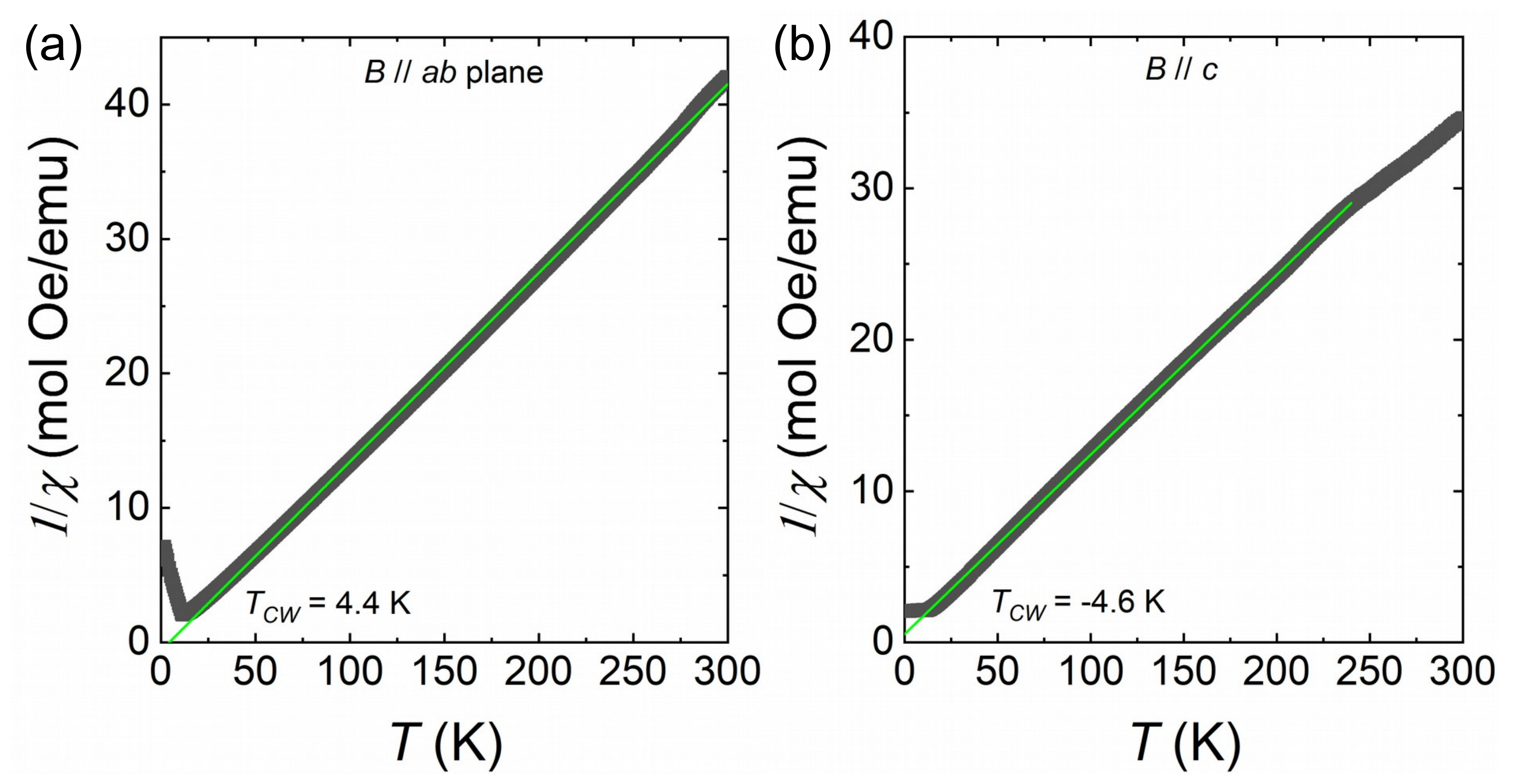}
\caption{\label{figS6}
Temperature dependence of the inverse susceptibility of EuAs$_3$ single crystal for magnetic fields applied (a) in the $ab$ plane, and (b) parallel to the $c$ axis. 
}
\end{figure*}

\begin{figure*}[h!]
\renewcommand\thefigure{S7}
\includegraphics[clip,width=1\textwidth]{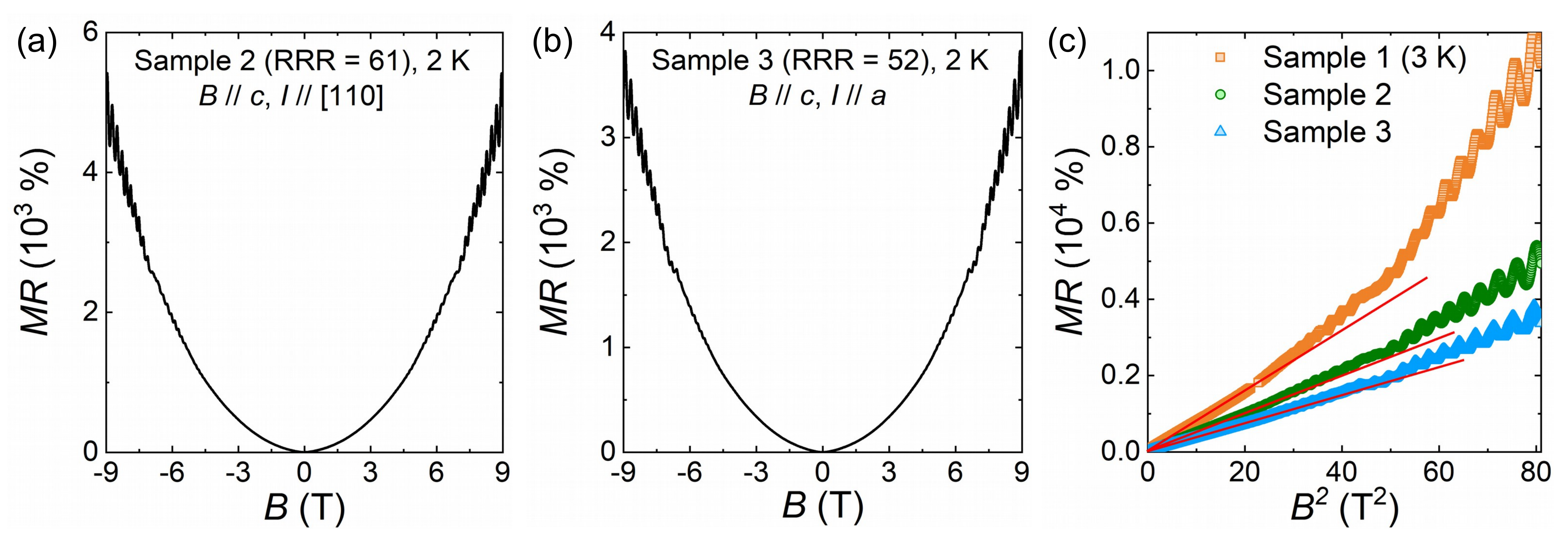}
\caption{\label{figS7}
Magnetoresistance (MR) measurements of EuAs$_3$ single crystal at 2 K with different current direction. (a) For Sample 2, magnetic field is applied along the $c$ axis, and electric current $I$ along the [110] direction. (b) For Sample 3, magnetic field is applied along the $c$ axis, and electric current $I$ along the $a$ axis. (c) $B^2$ as a function of MR for three different samples. The data for Sample 1 is taken from that in Fig.\ 2(b). With changing the direction of electric current, the unsaturated XMR persists, excluding the open-orbit effect \cite {XMR-2}. 
}
\end{figure*}

\end{document}